\documentstyle[twocolumn,prl,aps,epsfig]{revtex}

\begin{document}

\newcommand{\rar}{\rightarrow}
\newcommand{\rp}{\right)}
\newcommand{\lp}{\left(}
\newcommand{\Q}{{\vec q}}
\newcommand{\K}{{\vec k}}
\newcommand{\R}{{\vec r}}
\newcommand{\Rp}{{\vec r}^\prime}
\newcommand{\myO}{$\ddot{\mbox{o}}$}


\twocolumn[\hsize\textwidth\columnwidth\hsize\csname
@twocolumnfalse\endcsname

\title{Electronic properties of the degenerate Hubbard Model~: \\
A dynamical mean field approach}

\author{Pierre Lombardo, Gilbert Albinet}
\address{L2MP\\
Marseille, France}
\date{\today}
\maketitle

\begin{abstract}
We have investigated electronic properties of the degenerate
multi-orbital Hubbard model, in the limit of large spatial dimension.
A new local model, including a doubly degenerate strongly correlated site 
has been introduced and solved in the framework of the non-crossing
approximation (NCA). Mott-Hubbard transitions have been examined
in details, including the calculation of Coulomb repulsion critical values
and electronic densities of states for any regime of parameters.
\end{abstract}
\vspace{5mm}
]

\narrowtext

%
An increasing interest for strongly correlated electronic
systems has been induced by the discovery of high temperature
superconductors~\cite{BM86} and giant
magneto-resistance~\cite{HWH93}.
Despite a large amount of papers on this fascinating field,
a complete physical understanding of strongly correlated
electronic systems remains a major problem of
condensed matter theory. Even the simplest Hamiltonian
describing such systems, given by the Hubbard model,
is still an unsolved challenging problem.

The seminal work of Metzner and Vollhardt~\cite{MV89}
has shown the importance of large spatial dimensions $D$.
Taking the limit $D\longrightarrow\infty$ leads to significant
simplification of the many body problem while retaining essential
dynamical features of low dimensional situations.
An important amount of studies has been done within this
approach, mapping the lattice problem onto a self-consistent
Anderson single impurity model.
This mapping becomes exact in the
limit of infinite spatial dimension. For a review,
see~\cite{GKK96}.
Most of this work has been done focusing on the
original one-band Hubbard model, which is the simplest
model to describe the electron-electron interaction-driven
metal-insulator transition (MIT).

However, orbital degeneracy in known to play
a crucial role in correlated systems. The degeneracy
of the $d$ band is two in V$_2$O$_3$~\cite{CNR78}
and three in LaTiO$_3$.
Orbital degree of freedom is relevant to explain
some very interesting properties like colossal
magneto-resistance~\cite{MMS96},
MIT in alkali-doped fullerenes~\cite{W92,POK93},
and for any physical property involving orbital ordering.

Many theoretical approaches have been proposed to describe
the effect of strong Coulomb interaction in systems with
orbital degeneracy, using
the slave-boson method~\cite{L94,H97,FK97,KS98},
the variational method~\cite{BW97},
and the limit of high spatial dimension.
Some results of these works concerning the transition criteria
and the order of the MIT are substantially different.
An unified theory describing degenerate Hubbard model
is still missing. Concerning high spatial dimension
approaches, quantum Monte-Carlo (QMC)~\cite{R97,HJC98} and
a generalized iterated perturbation theory (IPT)~\cite{KK96}
have been proposed.
Imaginary time results, high computational time
and fundamental difficulties at low temperature
can be a limit for QMC calculations. The generalization of
the IPT, developed in Ref.~\cite{KKbis96}, presents the advantage
of being able to deal with particle-hole asymmetric problems.
However, this approach is essentially a perturbation theory
with respect to the correlation strength, which is the highest
energy scale of the system.
The non-crossing approximation (NCA)~\cite{B87} was shown to be
in excellent agreement with quantum Monte Carlo
calculations for the DMFT of the one-orbital case~\cite{JP93,PCJ93},
even far away from half-filling.

Here, we present a new approach based on the NCA within
the dynamical mean field theory (DMFT).

%
The starting Hamiltonian is given by the two-orbital degenerate Hubbard
model
\begin{eqnarray}
H=& &\sum_{<i,j>,\sigma,a,b}t^{ab}_{ij}c^+_{ia\sigma} c_{jb\sigma}
+\frac{U+J}{2}\sum_{i,a,\sigma} n_{ia\sigma}n_{ia-\sigma} \nonumber\\
&+& \frac{U}{2}\sum_{i,a\neq b,\sigma}n_{ia\sigma}n_{ib-\sigma}
+ \frac{U-J}{2}\sum_{i,a\neq b,\sigma}n_{ia\sigma}n_{ib\sigma} \nonumber\\
&-& \frac{J}{2}\sum_{i,a\neq b,\sigma}c^+_{ia\sigma}c_{ia-\sigma}
c^+_{ib-\sigma}c_{ib\sigma} \label{hamiltonien}
\end{eqnarray}
where the sum $<i,j>$ is the sum over nearest neighbor sites
of a Bethe lattice and $a,b=1,2$ is the band index.
$c^+_{ia\sigma}$ (respectively $c_{ia\sigma}$)
denotes the creation (respectively annihilation)
operator of an electron at the lattice site $i$ with spin
$\sigma$ and orbital index $a$ and  $n_{ia\sigma}$ is
the occupation number per spin and per orbital.
The on-site Coulomb repulsion $U$ and the
exchange parameter $J$ are assumed to be independent
of orbital. In addition we will neglect the last term in the
Hamiltonian~(\ref{hamiltonien}) and the hopping between different
orbitals $t^{ab}_{ij}=-t\delta_{ab}$. Doing such approximations
leads to the same Hamiltonian studied previously by quantum
Monte-Carlo (QMC)~\cite{R97}.

Integrating out the fermionic degrees of freedom leads to
the single-site effective action
\begin{eqnarray*}
&S&_{\mathrm eff}=-\int_{0}^{\beta}d\tau\int_{0}^{\beta}d\tau'\sum_{\sigma,a}
c^+_{oa\sigma}(\tau){\cal G}^{-1}_{oa}(\tau-\tau')c_{oa\sigma}(\tau')\\
&+& \frac{U}{2}\int_{0}^{\beta}d\tau\left(
\sum_{a\sigma}n_{oa\sigma}(\tau) n_{oa-\sigma}(\tau)
+\sum_{(a\neq b)}n_{oa}(\tau) n_{ob}(\tau)
\right)~,
\end{eqnarray*}
where
$$
n_{oa}=n_{oa\uparrow}+n_{oa\downarrow}~.
$$

\begin{figure}
 \epsfig{file=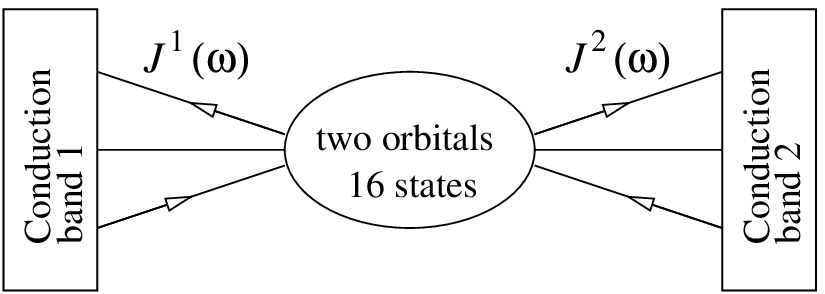, width=8.4cm}
 \caption{Two orbitals local model. The
 sixteen local states $|\alpha,\beta>$ are all coupled
 by the orbital dependent effective medium hybridization
 ${\cal J}^1(\omega)$ and ${\cal J}^2(\omega)$.}
 \label{modele}
\end{figure}

Therefore, as in the one-band case, the large-$D$ version of the
degenerate two-band Hubbard model is mapped onto an effective
impurity model. The local correlated site is now orbitally degenerated,
and two different effective media have to be considered.
The corresponding Hamiltonian reads
\begin{equation}
H_{\mathrm eff}=H_{\mathrm loc}+H_{\mathrm med}~,
\label{model}
\end{equation}
where the local part is
\begin{eqnarray*}
H_{\mathrm loc}&=&\sum_{a\sigma}\varepsilon_d n_{oa\sigma}\\
&+&\frac{U}{2}\left(
\sum_{a\sigma}n_{oa\sigma} n_{oa-\sigma}
+\sum_{(a\neq b)}n_{oa} n_{ob}
 \right)~,
\end{eqnarray*}
and the coupling with the effective medium is
\begin{eqnarray*}
H_{\mathrm med}&=&\sum_{\K a\sigma}
\left(W_\K^a b^+_{a\K \sigma} c_{a\sigma}+H.c.\right)\\
&+&\sum_{\K a\sigma}\varepsilon_\K^a b^+_{a\K \sigma}b_{a\K \sigma}~.
\end{eqnarray*}
The orbital dependent coupling with the effective medium is
characterized by the hybridization
$$
{\cal J}^a(\omega)=\sum_{\K}\frac{|W_\K^a|^2}{\omega+i0^+-\varepsilon_\K^a}~.
$$
From the equation of motion of this effective Hamiltonian, it follows that
$$
G_{a\sigma}(\omega)^{-1}=\omega-\varepsilon_d-\Sigma_{a\sigma}(\omega)
-{\cal J}^a(\omega)~.
$$
Comparing this equation to the following property of Green's functions
on a Bethe lattice
$$
G_{a\sigma}(\omega)^{-1}=\omega-\varepsilon_d-\Sigma_{a\sigma}(\omega)
-t^2 G_{a\sigma}(\omega)~,
$$
we found the self-consistent set of equations
$$
{\cal J}^a(\omega)=t^2 G_{a\sigma}(\omega)~.
$$
In the following, we solve the impurity model of equation~(\ref{model})
using the extended version of the non-crossing approximation presented
in a previous work~\cite{LAS96}. Propagators and self-energies of the
sixteen local states $|\alpha,\beta>$ are introduced,
where $\alpha$ (respectively $\beta$) is the local occupation of
orbital $a=1$ (respectively $a=2$). Each local state $|\alpha,\beta>$
is an eigenstate of the local part of the model Hamiltonian.
The corresponding local model that have to be solved is 
schematically represented in figure~\ref{modele}. The impurity
site is coupled with two effective media by two effective 
dynamical hybridization ${\cal J}^1(\omega)$ and ${\cal J}^2(\omega)$.

Local states $|\alpha,\beta>$ and $|\alpha',\beta'>$ are coupled
by NCA equations if one of the two following conditions
are fulfilled~:$|n_{o1}(\alpha)-n_{o1}(\alpha')|=1$ and $\beta=\beta'$
or $\alpha=\alpha'$ and $|n_{o2}(\beta)-n_{o2}(\beta')|=1$. The first 
and the second cases involve respectively 
${\cal J}^1(\omega)$ and ${\cal J}^2(\omega)$.

%

%

Here are the main results of our dynamical mean field theory.
The NCA approach presents the advantage to provide directly
real frequencies one-particle Green's functions.
Local quantities like densities of states can therefore be computed
for any regime of parameters. 
\begin{figure}[h]
 \epsfig{file=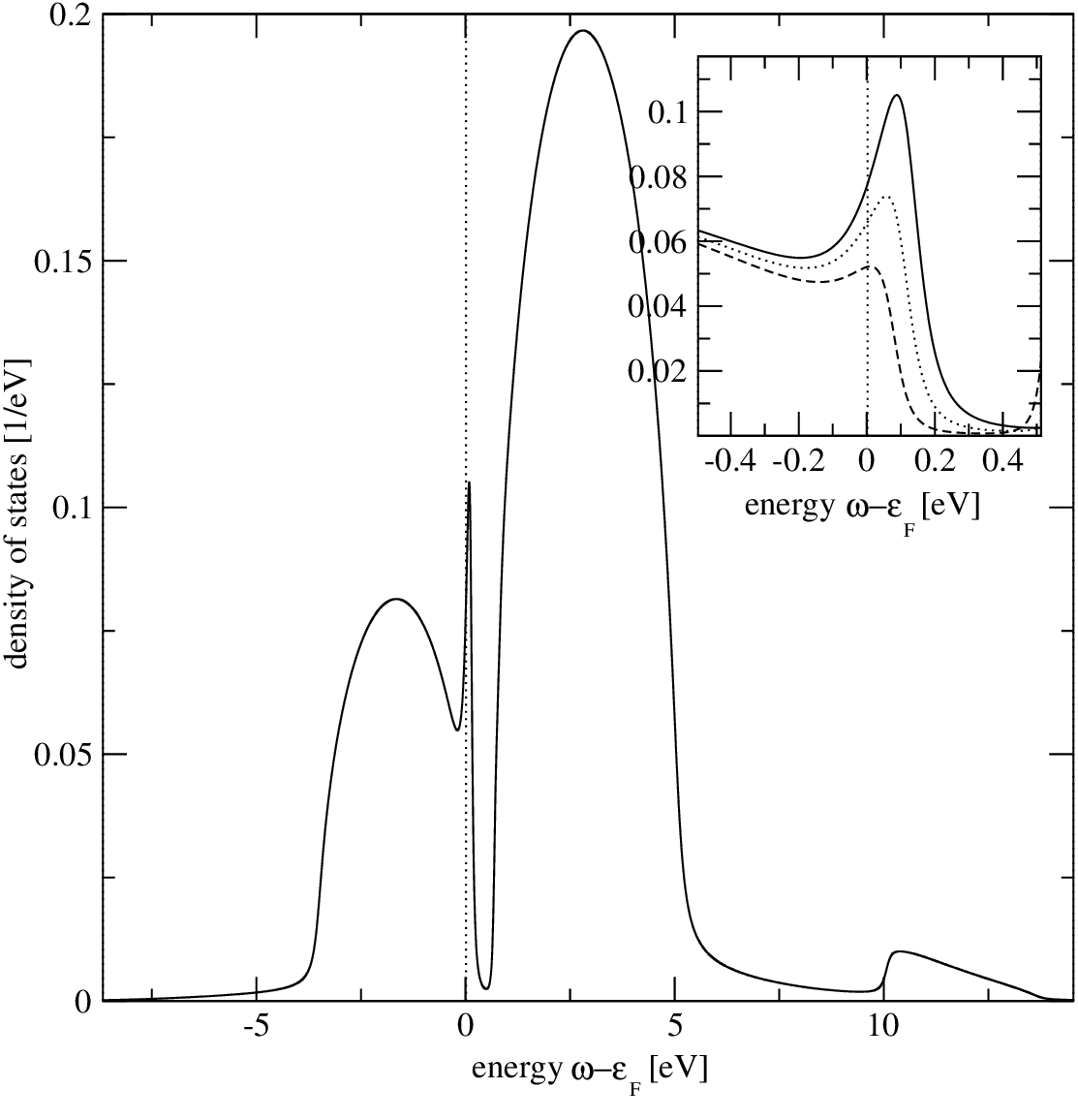, width=8.4cm}
 \caption{Density of states per spin and per orbital for $U=4$~eV,
 $t=1$~eV and $T=1000$~K. Total occupation number is $n=0.987$.
 The inset shows filling dependence of low energy excitations.
 Solid line (respectively dotted and dashed line) is the density
 of states around the Fermi level for $n=0.987$
 (respectively $n=0.992$ and $n=0.995$).}
 \label{figdos0}
\end{figure}

\noindent NCA was in very good agreement with
essentially exact quantum Monte-Carlo simulations for the one-band
case. This approximation is then a good candidate for the
investigation of the multi-band generalization of the Hubbard model.

Useful and comprehensive information
concerning electronic dynamic processes are given by these
densities of states, without any size effects due to small clusters
and till very low temperature.

Spin and orbital dependent densities of states $\rho_{a\sigma}(\omega)$
are obtained from NCA Green's functions by~:
$$
\rho_{a\sigma}(\omega)=-\frac{1}{\pi}\rm{Im}\{ G_{a\sigma}(\omega) \}~.
$$

Figure~\ref{figdos0} displays the density of states per spin and per orbital
of the doubly degenerate Hubbard model for a total occupation number
$n=0.987$. Both lower and upper Hubbard bands are present.
Relative spectral weights between the lower and the upper Hubbard band
are given by the ratio $1/3$.
Quasiparticle excitations can be seen around the Fermi level.
The origin of this low energy structure can be clarified
in the context of NCA by observing its temperature and
filling dependence~\cite{PCJ93} or by 
\begin{figure}[h]
 \epsfig{file=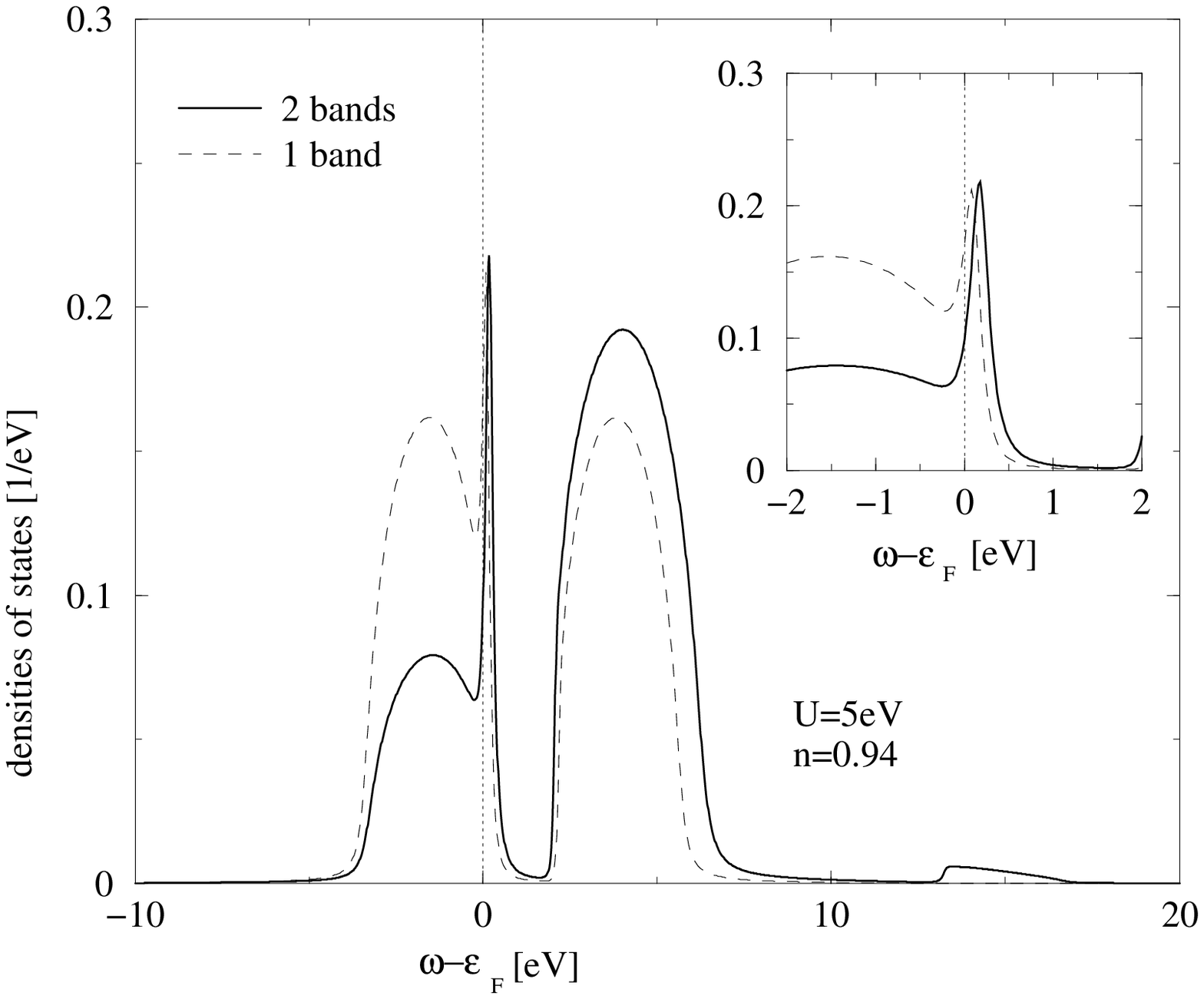, width=8.4cm}
 \caption{Comparison of the densities of states per spin
 and per orbital for $U=5$~eV, $t=1$~eV and $T=1000$~K
 in the one-orbital case (dashed line) and the two-orbitals case (solid line)
 Total occupation number is $n=0.94$ in both cases.
 The inset shows low energy excitations in both cases
 for the same parameters.}
 \label{fig1b2b}
\end{figure}

\noindent reducing the Hilbert 
space of available local states onto the impurity~\cite{LAS96}.

The proximity of the $n=1$ metal insulator transition is confirmed
by the disappearance of the low energy spectral weight. The inset 
shows the behavior of quasiparticle excitations just before
the transition. Total occupation numbers are $n=0.987$, $n=0.992$ and
$n=0.995$.
%

%

It is interesting to compare two-bands densities of states with
one-band NCA results. This is shown in figure~\ref{fig1b2b},
for $U=5$~eV and $n_{tot}=0.94$. Spectral weights are in good
agreement with previous approaches like the IPT one~\cite{KK96}.
More precisely, we found that degeneracy has a strong influence
only on high energy excitations~: respective spectral weights
of lower and upper Hubbard bands for $n_{tot}=1$ are $1$ and
$3$ in the two-orbitals problem, instead of $1$ and $1$ for
the same filling in the one-orbital problem.
However, NCA bands structures are strongly different to IPT ones.
Fine structures displayed by IPT results are not present in
our approach which is confirmed by
recently obtained QMC spectral functions for the multi-orbital
Hubbard model in infinite dimension~\cite{HJC98}.
%

%

\begin{figure}[h]
 \epsfig{file=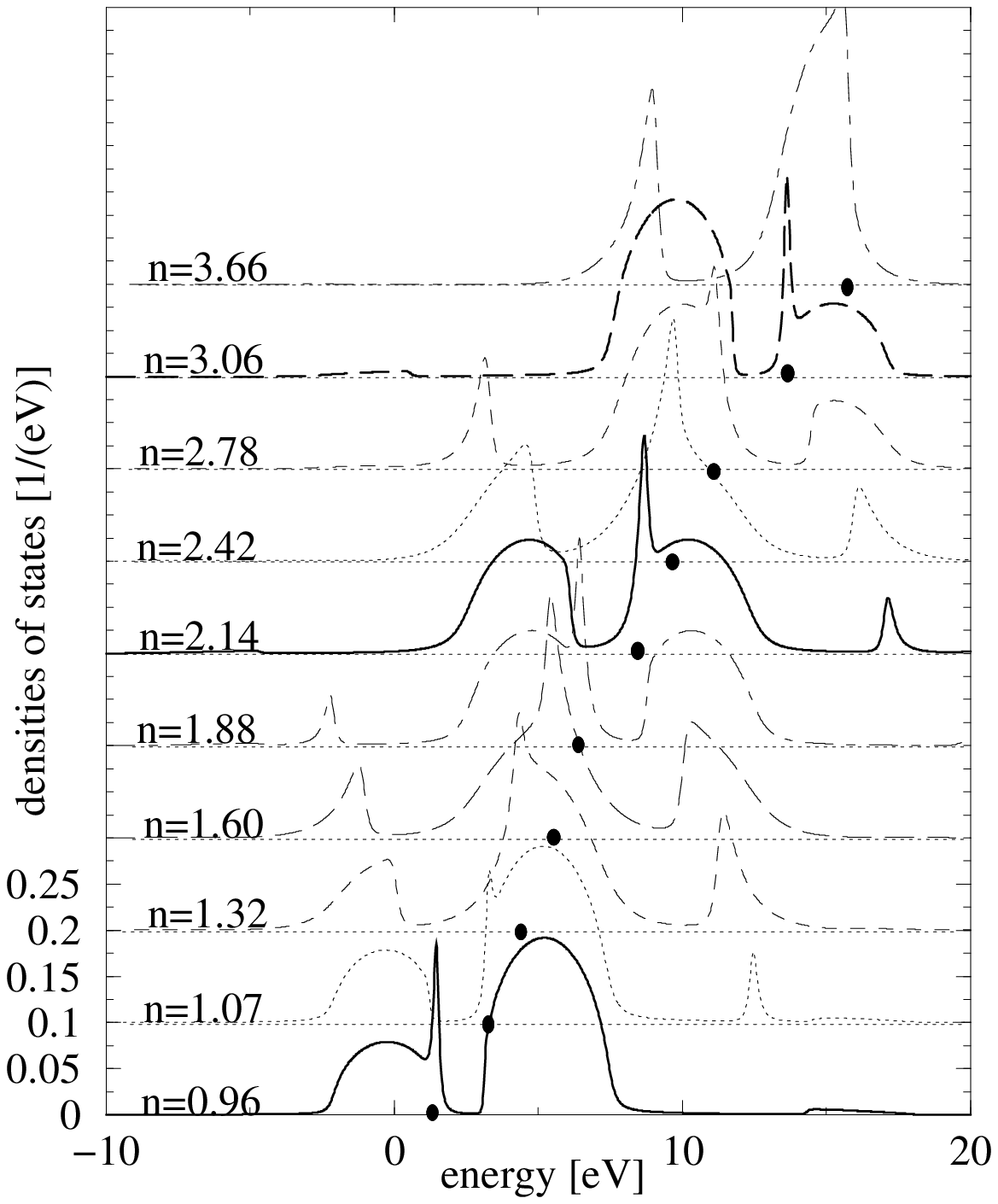, width=8.4cm}
 \caption{Densities of states per spin and per orbital for
 various fillings. The correlation strength is $U=5$~eV, bandwidth
 is $t=1$~eV and temperature is $T=1000$~K. For each curve,
 Fermi level is labeled by a black point.}
 \label{figdos}
\end{figure}

Figure~\ref{figdos} shows the general evolution with respect to
the total occupation number of
electron density of states per spin and per orbital,
for $U=5$~eV, $t=1$~eV and $T=1000$~K. Each curve
corresponds to a given Fermi level. Quasiparticle low energy excitations
can be seen around the Fermi level. Total occupation numbers $n$ are indicated
for each density of states. For integer fillings $n=1$, $n=2$ and $n=3$,
the system is an insulator. It is metallic for any other filling.
The interesting transfer of spectral weight from high to low (Fermi level)
energy scale will be studied in details later.

Note that effective mass of quasiparticles increases when approaching the Mott
insulating state, which is consistent with experimental observations in filling
controlled oxides family such as Sr$_{1-x}$La$_{x}$TiO$_3$~\cite{TTO93}.
%

\begin{figure}
 \epsfig{file=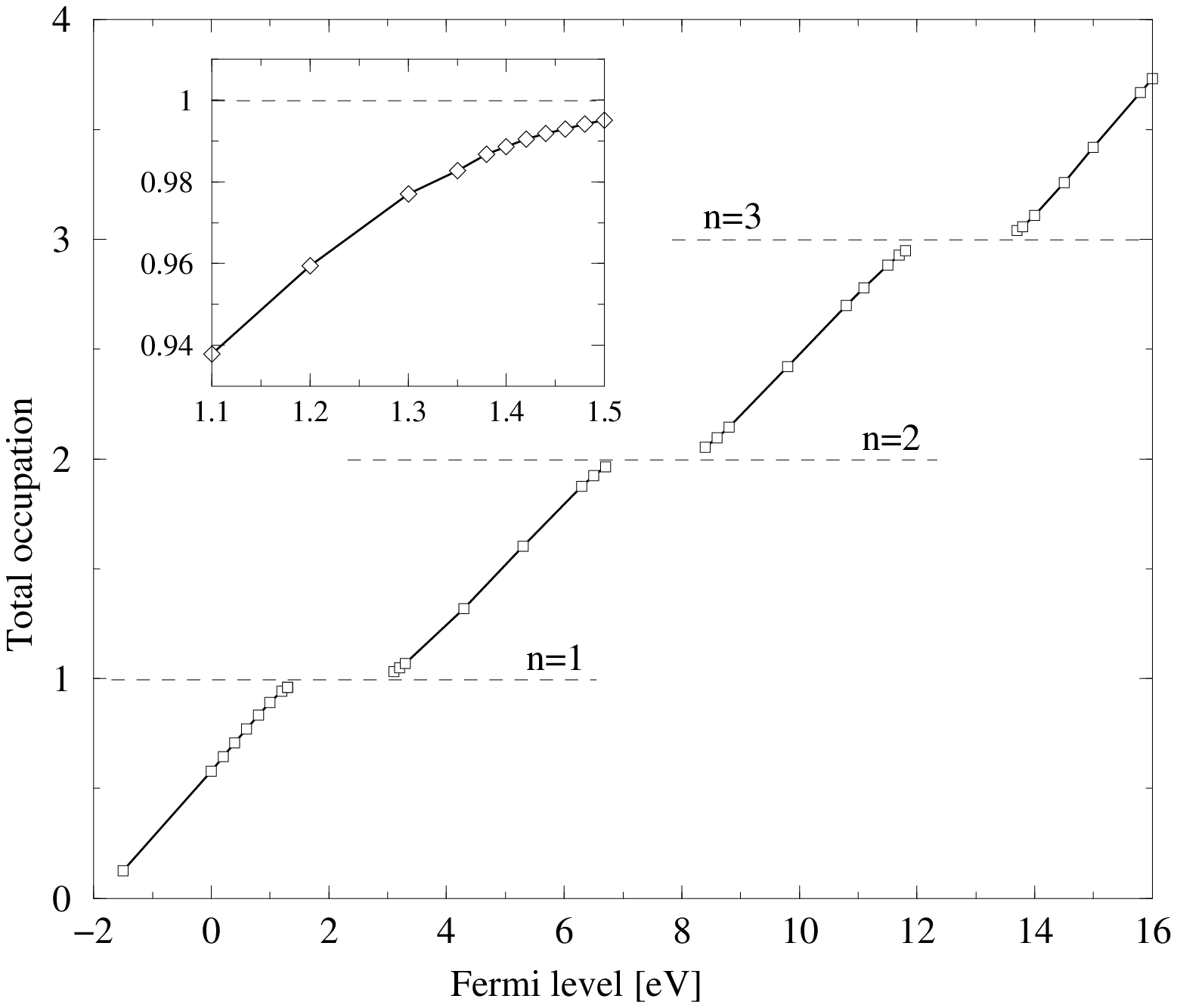, width=8.4cm}
 \caption{Total occupation number with respect to
 Fermi level of $d$ electrons. Parameters are $U=5$~eV, $t=1$~eV
 and temperature is $T=1000$~K. The inset shows
 the occupation up to the transition value $n=1$, for
 $U=4$~eV.}
 \label{figpalliersMIT}
\end{figure}

The Fermi level dependence of total occupancy is plotted in
figure~\ref{figpalliersMIT}. These results are consistent
with recent QMC simulations for the multi-orbital
Hubbard model~\cite{HJC98}. When the system in an insulator,
as for $U=4$~eV in the figure, a small displacement of the Fermi
level in the insulating gap does not lead to a significant
variation of the total occupation number. A plot of occupation
number versus chemical potential displays a plateau as soon as
a MIT is crossed. Our results show that three successive
MIT at integer fillings occur in the system.
$n=1$ and $n=3$ transitions are equivalent because of the
particle-hole symmetry. For a given $U$, the Mott gap
of the $n=2$ transition is smaller than the $n=1$ gap.
This is in good agreement with previous calculations
leading to a larger critical $U_c$ for the $n=2$ transition~\cite{R97}.
We found $U_c(n=2)\approx 3.5$ and $U_c(n=1)\approx 3.1$.
%

%

%
In summary, we have proposed a new approach based
on the dynamical mean field
theory for the doubly degenerate Hubbard model, combined with a
generalization of the non-crossing approximation in order to
solve the equivalent self-consistent local impurity problem.
This impurity problem consists in a doubly degenerate
local correlated site (sixteen local states) embedded
in a conduction band which is orbital dependent.

We obtained a description of the integer filling metal insulator transitions
consistent with previous theoretical works using slave-boson formalism,
variational method and quantum Monte-Carlo dynamical mean field theory.
One of the advantages of our approach is to provide
the full spectrum of excitations for any regime
of parameters and a clear interpretation of the
quasiparticle nature. Besides, many extensions of the model Hamiltonian can
be done within this framework. In future calculations, we will
increase the degeneracy (some oxides, like Sr$_{1-x}$La$_{x}$TiO$_3$
are triply degenerate) and we will take explicitly into account
the exchange interaction by introducing a non zero Hund coupling
parameter $J$. We will also extend our analysis to
thermodynamic and transport properties.

\end{document}